\documentclass[twocolumn]{pasj01}
\usepackage{natbib,bm,url,color}
\newcommand{\pbf}{}

\begin{document} 
\Received{}
\Accepted{}

\title{Formation rate of LB-1-like systems through dynamical interactions}

\author{Ataru \textsc{Tanikawa}\altaffilmark{1}}%
\altaffiltext{1}{Department of Earth Science and Astronomy, College of
  Arts and Sciences, The University of Tokyo, 3-8-1 Komaba, Meguro-ku,
  Tokyo 153-8902, Japan}
\email{tanikawa@ea.c.u-tokyo.ac.jp}

\author{Tomoya \textsc{Kinugawa}\altaffilmark{2,3}}
\altaffiltext{2}{Department of Astronomy, Graduate School of Science,
  The University of Tokyo, 7-3-1 Hongo, Bunkyo-ku, Tokyo 113-0033,
  Japan}
\altaffiltext{3}{Institute for Cosmic Ray Research, The University of
  Tokyo, Kashiwa, Chiba 277-8582, Japan}

\author{Jun \textsc{Kumamoto}\altaffilmark{2}}

\author{Michiko S. \textsc{Fujii}\altaffilmark{2}}


\maketitle

\begin{abstract}

We estimate formation rates of LB-1-like systems through dynamical
interactions in the framework of the theory of stellar evolution
before the discovery of the LB-1 system. The LB-1 system contains
$\sim 70M_\odot$ black hole (BH), so-called pair instability (PI)-gap
BH, and B-type star with solar metallicity, and has nearly zero
eccentricity. The most efficient formation mechanism is as follows. In
an open cluster, a naked helium (He) star (with $\sim 20M_\odot$)
collides with a heavy main-sequence (MS) star (with $\sim 50M_\odot$)
which has a B-type companion. The collision results in a binary
consisting of the collision product and B-type star with a high
eccentricity. The binary can be circularized through the dynamical
tide with radiative damping of the collision-product
envelope. Finally, the collision product collapses to a PI-gap BH,
avoiding pulsational pair instability and pair instability supernovae
because its He core is as massive as the pre-colliding naked He
star. We find that the number of LB-1-like systems in the Milky Way
galaxy is $\sim {\pbf 0.01} (\rho_{\rm oc} / 10^4 M_\odot
\mbox{pc}^{-3})$, where $\rho_{\rm oc}$ is the initial mass densities
of open clusters. If we take into account LB-1-like systems with
O-type companion stars, the number increases to $\sim {\pbf 0.03}
(\rho_{\rm oc} / 10^4 M_\odot \mbox{pc}^{-3})$. This mechanism can
form LB-1-like systems at least {\pbf 10} times more efficiently
than the other mechanisms: captures of B-type stars by PI-gap BHs,
stellar collisions between other type stars, and stellar mergers in
hierarchical triple systems. {\pbf We conclude that no dynamical
  mechanism can explain the presence of the LB-1 system.}

\end{abstract}

\section{Introduction}


Stellar-mass black holes (BHs) are the end state of massive
stars. They have long been observed only as X-ray binaries
\citep[][for review]{2006ARA&A..44...49R}. In the last few years,
however, gravitational wave radiations have been successfully detected
from mergers of binary BHs (BH-BHs)
\citep{2016PhRvL.116f1102A,2016PhRvL.116x1103A,2017PhRvL.118v1101A,2017PhRvL.119n1101A,2017ApJ...851L..35A,2019PhRvX...9c1040A,2019arXiv190407214V,2019arXiv191009528Z}. Since
these achievements accelerate BH explorations, spectroscopic
observations have recently discovered BHs in wide binaries, i.e. those
without interactions with their luminous companion stars
\citep{2018ApJ...856..158K,2018MNRAS.475L..15G,2019Sci...366..637T,2019Natur.575..618L}. These
discoveries are rapidly putting forward our understanding of BHs.

The LB-1 system, a binary composed of a BH and luminous companion
star, has been discovered by \cite{2019Natur.575..618L} using
spectroscopic observations. The BH mass of the LB-1 system was
estimated to be $\sim 70M_\odot$, and the luminous companion has solar
(or supersolar) metallicity. The estimated binary eccentricity $\sim
0.03$, nearly zero. The formation of LB-1-like system is quite
challenging for the currently known theories of single stellar
evolution for the following reasons. Massive stars evolve to naked
helium (He) stars due to strong stellar-wind mass-loss under solar
metallicity environment, and as a result they leave at most
$20M_\odot$ BHs
\citep[e.g.][]{2001A&A...369..574V,2010ApJ...714.1217B}. Even if
stellar wind does not work well for some reason, massive stars with
large He cores should undergo pair instability supernovae (PISNe)
{\pbf
  \citep{1967PhRvL..18..379B,1968Ap&SS...2...96F,1984ApJ...280..825B,1986A&A...167..274E,2001ApJ...550..372F,2002ApJ...567..532H,2002ApJ...565..385U}}
or pulsational pair instability supernovae (PPISNe) {\pbf
  \citep{2002ApJ...567..532H,2007Natur.450..390W,2016MNRAS.457..351Y,2017ApJ...836..244W,2019ApJ...887...72L}.}
These effects limit BH masses to less than $\sim 50M_\odot$ under any
metallicity environments \citep{2016A&A...594A..97B}. {\pbf
  Metal-poor or Pop.~III stars} with a mass of $\gtrsim 300M_\odot$
first overcome PPISN/PISN effects, and directly collapse to BHs with
little mass loss \citep{2002ApJ...567..532H}. Therefore, there should
be no BH in a mass range from $\sim 50M_\odot$ to $\sim 300M_\odot$ if
we consider single stellar evolution theories, and this mass range is
called pair instability (PI) gap \citep{2019ApJ...882L..24A}. In spite
of these theoretical predictions, the LB-1 system has a BH with $\sim
70M_\odot$. In response to this discovery, several studies have
reconsidered massive star evolution; they have reduced effects of
stellar wind mass loss
\citep[e.g.][]{2019arXiv191112357B,2019arXiv191200994G}. {\pbf
  Note that several studies have raised doubts on the presence of the
  $70M_\odot$ BH
  \citep{eldridge2019weighing,abdulmasih2019signature,2019arXiv191210456S,2020MNRAS.493L..22E,2020A&A...633L...5I}.}

BH-BH mergers to form $\sim 70M_\odot$ BH were detected by
gravitational wave (GW) detectors such as LIGO and Virgo
\citep[e.g.][]{2016PhRvL.116f1102A,2019PhRvX...9c1040A}. If an inner
BH-BH of a hierarchical triple system with a B-type third star merged,
LB-1-like systems seem to be easily formed. However, this process
would leave a LB-1-like system with a high eccentricity (hereafter,
eccentric LB-1-like system), since the merged BH receives a kick with
a velocity of $\gtrsim 100$~km~s$^{-1}$ due to the asymmetric GW
radiation of the BH-BH merger
\citep[e.g.][]{2007PhRvD..76f4034B,2007PhRvL..98w1102C,2010CQGra..27k4006L}. This
binary could be circularized by tidal interaction, but the time scale
is estimated to exceed the Hubble time
\citep{2019Natur.575..618L}. {\pbf If the $70M_\odot$ BH is not a
  single BH, but unresolved double BHs,} we can avoid such a high kick
velocity, but several studies have already ruled out this possibility
\citep{2019Natur.575..618L,2019arXiv191112581S}.

Another scenario is capturing a B-type star after a BH-BH merger. If
two BHs merge in a globular cluster, the merged BH can be retained in
the globular cluster \citep{2019PhRvD.100d3027R}. The merged BH could
capture a B-type star through a binary-single encounter. However, such
an encounter should also form a binary with a large eccentricity
because the eccentricity of dynamically formed binaries has a thermal
distribution \citep{1975MNRAS.173..729H}.

\begin{figure*}
 \begin{center}
  \includegraphics[width=120mm]{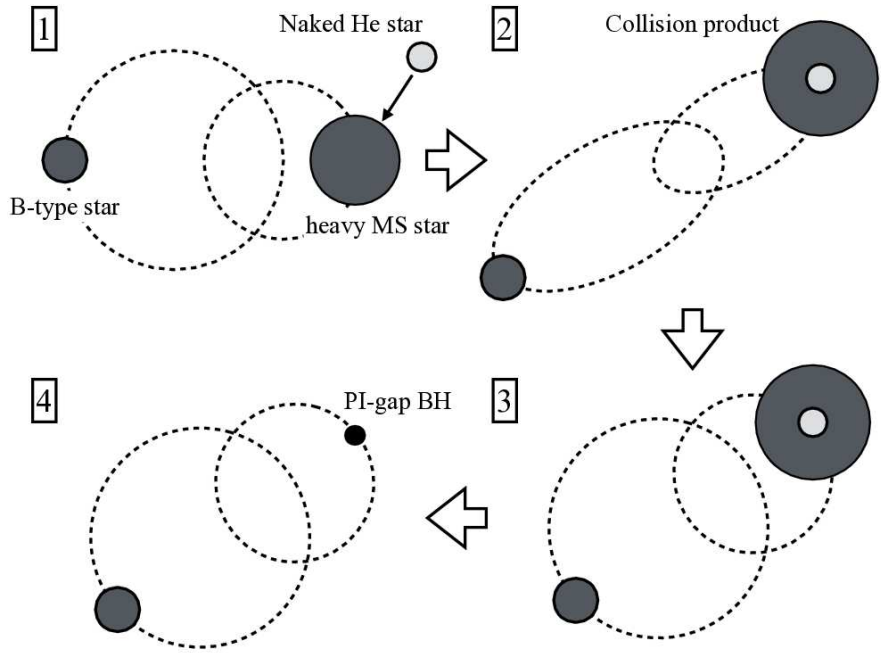} 
 \end{center}
 \caption{Schematic figure of the formation process of LB-1
   system.}\label{fig:illustration}
\end{figure*}

In this paper, we assess a formation mechanism of LB-1-like systems in
open clusters. Some recent studies have suggested that open clusters
are one of probable formation sites of BH binaries
\citep[e.g.][]{2019MNRAS.486.3942K,2019MNRAS.487.2947D,2020arXiv200110690K,2020arXiv200111199S}. Here,
we do not assume the reduction of stellar-wind mass-loss, but
dynamical formation mechanisms. The formation mechanism we propose in
this paper is as follows (see also Figure \ref{fig:illustration}). We
consider a binary-star system with heavy main-sequence (MS) star with
$\sim 50M_\odot$ and B-type star with $\sim 8M_\odot$, and a single
naked He star with $\sim 20M_\odot$ in an open cluster. They form a
LB-1-like system through the following four processes. (The numbers of
these processes correspond to those in Figure \ref{fig:illustration}.)
\begin{enumerate}
    \item The binary and naked He star experience a close
      binary-single encounter. During this close encounter, the naked
      He star collides with the heavy MS star.
    \item This collision forms a He star with a massive hydrogen (H)
      envelope. The collision product and B-type star compose a new
      binary. This binary has a finite eccentricity due to the
      collision.
    \item The envelope of the collision product exerts tidal friction
      on the binary motion. Then, the binary orbit is circularized.
    \item The collision product directly collapses to a PI-gap BH with
      a mass of $\sim 70M_\odot$. Note that this collapse avoids PPISN
      and PISN, since the He core mass is $\sim 20M_\odot$.
\end{enumerate}
This binary finally escapes from the open cluster due to the two-body
relaxation process or due to a close encounter with another star
before the collision product collapses to a PI-gap BH. The formation
mechanism of PI-gap BHs via stellar collisions in open cluster have
been suggested in \cite{2019MNRAS.485..889S} and
\cite{2019arXiv191101434D}. However, no quantitative estimation for
the formation and circularization of binaries has been given
yet. {\pbf \cite{2019arXiv190808775O} have presented the
  synthetic catalog of BHs in the Milky Way (MW) galaxy, however the
  catalog has contained BHs formed in the MW galactic field, not BHs
  formed in the MW open clusters.}

The collision product and B-type star are always bound because this
binary-single encounter has negative energy owing to small velocity
dispersion ($\sim 1$ km s$^{-1}$) in open clusters. Note that the
collision never yields some additional energy, differently from
collisions of two white dwarfs
\citep{2009MNRAS.399L.156R,2009ApJ...705L.128R,2010MNRAS.406.2749L,2015MNRAS.454L..61D}. \cite{2010MNRAS.402..105G}
have shown that binary-single encounters end up with mergers among
three stars. However, pre-existing binaries in their simulations are
relatively close: semi-major axes of several $10R_\odot$. On the other
hand, we suppose pre-existing binaries with semi-major axes of
$1$~au. Thus, three stars do not necessarily merge in our cases.
After the formation, these binaries would never keep staying in open
clusters. They are frequently ejected from open clusters through close
encounters \citep{2011Sci...334.1380F,2012ApJ...746...15B}. We can
estimate the separation between these binaries and their formation
sites as $\sim 1 (v_{\rm ej}/30\mbox{kms}^{-1}) (T_{\rm
  B}/40\mbox{Myr})$ kpc, where $v_{\rm ej}$ is the ejection velocity
of these binaries from open clusters, and $T_{\rm B}$ is the lifetime
of B-type stars. Therefore, the current locations of these binaries
are not necessarily close to open clusters.

The following is the structure of this paper. In section
\ref{sec:FormationRateOfLB-1-likeProgenitors}, we roughly count the
formation rate of binaries consisting of the collision products and
B-type stars in the MW galaxy. We call these binaries ``LB-1-like
progenitors''. In section \ref{sec:TidalCircularization}, we calculate
circularization timescale of the progenitors, and estimate the number
of LB-1-like systems in the MW galaxy. In section
\ref{sec:OtherPossibleScenarios}, we rule out any other scenarios
related to dynamical interactions. In section \ref{sec:Summary}, we
summarize this paper.

\section{Formation rate of LB-1-like progenitors}
\label{sec:FormationRateOfLB-1-likeProgenitors}

In this section, we estimate the formation rate of LB-1-like
progenitors formed in open clusters in the MW galaxy, $\dot{N}_{\rm
  LB1,p}$. We first define $\dot{N}_{\rm LB1,p}$ as
\begin{eqnarray}
    \dot{N}_{\rm LB1,p} &= \dot{N}_{\rm PIgap} \frac{\Gamma_{\rm
        nHe}}{\Gamma_{\rm eHe}} {\pbf P_{\rm
        b}}, \label{eq:FormationRateOfLB-1-likeProgenitors}
\end{eqnarray}
where $\dot{N}_{\rm PIgap}$ is the formation rate of PI-gap BHs in all
the MW open clusters, and $\Gamma_{\rm nHe}$ and $\Gamma_{\rm eHe}$
are collision rates between heavy MS and naked He stars, and between
heavy MS and He stars with H envelopes ({\pbf enveloped
  He-burning stars}, hereafter), respectively. PI-gap BHs do not
always become members of LB-1-like systems. PI-gap progenitors should
be formed through collisions between heavy MS and {\pbf naked He
  or enveloped He-burning stars.} However, our scenario works only for
collision between heavy MS stars and naked He stars as described in
section \ref{sec:OtherPossibleScenarios}. Thus, we need the factor
$\Gamma_{\rm nHe}/\Gamma_{\rm eHe}$. {\pbf We set $P_{\rm b}$ to
    a probability that either of naked He stars and heavy MS stars
    have B-type companions, and their separations are $\sim 1$~au.}

We first estimate $\dot{N}_{\rm PIgap}$, which can be expressed as
\begin{eqnarray}
    \dot{N}_{\rm PIgap} = f_{\rm PIgap} \eta_{m_{\rm c}} f_{\rm oc} \dot{M}_{\rm mw},
\end{eqnarray}
where $f_{\rm PIgap}$ is the number fraction of PI-gap BHs to an open
clusters, $\eta_{m_{\rm c}}$ is the expected number of zero-age MS
(ZAMS) stars heavier than a certain mass $m_{\rm c}(M_\odot)$ per
stellar mass, $f_{\rm oc}$ is the mass fraction of stars formed in
open clusters, and $\dot{M}_{\rm mw}$ is the star formation rate in
the MW galaxy. If we choose appropriate $m_{\rm c}$, we find that
$\eta_{m_{\rm c}} f_{\rm oc}\dot{M}_{\rm mw}$ is the BH formation rate
in all the MW open clusters. Using the results of
\citet{2019arXiv191101434D}, we obtain the formation rate as
\begin{eqnarray}
    \dot{N}_{\rm PIgap} &\sim& 2 \times 10^{-6} \left( \frac{f_{\rm
        PIgap}}{0.002} \right) \left( \frac{\rho_{\rm oc}}{10^4M_\odot
      \mbox{pc}^{-3}} \right) \nonumber \\
    &\times& \left( \frac{\eta_{20}}{0.003M_\odot^{-1}} \right) \left(
    \frac{f_{\rm oc}}{0.2} \right) \left( \frac{\dot{M}_{\rm
        mw}}{2M_\odot\mbox{yr}^{-1}} \right)
         [\mbox{yr}^{-1}]. \label{eq:FormationRateOfPiGap}
\end{eqnarray}
\cite{2019arXiv191101434D} have shown that $f_{\rm PIgap}=0.002$ for
solar metallicity in open clusters with the initial mass densities
$\rho_{\rm oc} \sim 10^4M_\odot \mbox{pc}^{-3}$. Note that $f_{\rm
  PIgap}$ should be proportional to $\rho_{\rm oc}$, since PI-gap BHs
are formed through collisions between heavy MS and {\pbf naked He
  or enveloped He-burning stars.} We obtain from the SSE code
\citep{Hurley+2000} that MS stars leave BHs when their masses are
$\gtrsim 20M_\odot$, {\pbf where we adopt the models of
  \cite{Belczynski+2010} and \cite{2002ApJ...572..407B} for the
  prescriptions of stellar wind and supernova, respectively}. Then, we
find $\eta_{20}=0.003M_\odot^{-1}$, assuming stellar initial mass
function (IMF) as the Kroupa IMF in the mass range from $0.08M_\odot$
to $150M_\odot$ \citep{2001MNRAS.322..231K}. We derive $f_{\rm oc}$ as
follows. The total mass of open clusters with a mass of more than
$10^3M_{\odot}$ born within 100\,Myr in 1\,kpc from the sun is
$5.3\times 10^4M_{\odot}$ \citep{2007A&A...468..151P}. From this, we
estimate that the current star formation rate density of stars born in
open clusters is $5.3\times10^{-4}M_{\odot}\,{\rm yr}^{-1}\,{\rm
  kpc}^{-2}$. Since the star formation rate density at a Galactic
radius of 8\,kpc is estimated to be $\sim
3\times10^{-3}M_{\odot}\,{\rm yr}^{-1}\,{\rm kpc}^{-2}$
\citep{2006A&A...459..113M}, we estimate that $f_{\rm oc}\sim
0.18$. The current total star-formation rate of the MW galaxy is
$\dot{M_{\rm MW}}=1.65\pm 0.19 M_{\odot}{\rm yr}^{-1}$
\citep{2016ARA&A..54..529B}.

Before calculating $\Gamma_{\rm nHe}/\Gamma_{\rm eHe}$, we give
formulae for any rates of stellar collision/encounter between type-1
and type-2 stars, $\Gamma$, such that
\begin{eqnarray}
    \Gamma = N_1 n_2 \sigma_{12} v_{12}, \label{eq:AnyEventRate}
\end{eqnarray}
where $N_1$ is the number of type-1 stars, $n_2$ is the number density
of type-2 stars, $\sigma_{12}$ is the cross section of the
collision/encounter, and $v_{12}$ is the relative velocity of these
stars. Note that type-1 and type-2 stars are interchangeable. The
cross section can be written as
\begin{eqnarray}
    \sigma_{12} = \pi R_{12}^2 \left( 1+
    \frac{2GM_{12}}{R_{12}v_{12}^2} \right), \label{eq:CrossSection}
\end{eqnarray}
where $G$ is the gravitational constant, $M_{12}$ is the total mass of
type-1 and type-2 stars, and $R_{12}$ is critical separation between
type-1 and type-2 stars below which these stars are considered to
collide/encounter. Usually, the second term in the parentheses is
dominant. Thus, we define the product of $\sigma_{12}$ and $v_{12}$ as
a sweeping volume per unit time, $V_{12}$:
\begin{eqnarray}
    V_{12} \equiv \sigma_{12} v_{12} \sim 2 \pi
    GM_{12}R_{12}v_{12}^{-1}. \label{eq:SweepingVolume}
\end{eqnarray}

Now, we can express $\Gamma_{\rm nHe}/\Gamma_{\rm eHe}$ as
\begin{eqnarray}
    \frac{\Gamma_{\rm nHe}}{\Gamma_{\rm eHe}} &=& \frac{N_{1,{\rm
          nHe}}n_{2,{\rm nHe}}M_{12,{\rm nHe}}R_{12,{\rm
          nHe}}v_{12,{\rm nHe}}^{-1}}{N_{1,{\rm eHe}}n_{2,{\rm
          eHe}}M_{12,{\rm eHe}}R_{12,{\rm eHe}}v_{12,{\rm eHe}}^{-1}}
    \nonumber \\
    &=& \frac{N_{1,{\rm nHe}}M_{12,{\rm nHe}}R_{12,{\rm
          nHe}}}{N_{1,{\rm eHe}}M_{12,{\rm eHe}}R_{12,{\rm eHe}}},
\end{eqnarray}
where we add the subscripts "${\rm nHe}$" and "${\rm eHe}$" to all the
variables for the naked-He and {\pbf enveloped He-burning} cases,
respectively. We can obtain the second equality, considering that both
of $n_{2,{\rm nHe}}$ and $n_{2,{\rm eHe}}$ are the number density of
heavy MS stars, and both of $v_{12,{\rm nHe}}$ and $v_{12,{\rm eHe}}$
are velocity dispersion in open clusters. The number ratio of
$N_{1,{\rm nHe}}/N_{1,{\rm eHe}}$ can be interpreted as the ratio of
lifetimes of naked He and {\pbf enveloped He-burning} stars,
which should be $\sim 2$ in solar metallicity {\pbf from the SSE
  code with the stellar wind and supernova models described above}. We
set $M_{12,{\rm nHe}}/M_{12,{\rm eHe}} \sim 0.7$, supposing that naked
He stars have $\sim 20M_\odot$, and {\pbf enveloped He-burning}
stars and heavy MS stars have $\sim 50M_\odot$. {\pbf The masses
  of the enveloped He-burning stars can range from $\sim 20M_\odot$ to
  $\gtrsim 50M_\odot$ for the following reason. The ZAMS masses of the
  enveloped He-burning stars should be larger than $50M_\odot$, since
  their evolution is more rapid than the heavy MS stars. They reduce
  their masses due to stellar wind mass loss, and can be nearly naked
  He stars. Therefore, we choose $\sim 50M_\odot$ for their masses as
  a representative value.} Although $R_{12,{\rm nHe}}$ and $R_{12,{\rm
    eHe}}$ are the sum of the radii of two colliding stars, one of two
stars has a much larger radius than the other for both of the naked-He
and {\pbf enveloped He-burning} cases. Thus, $R_{12,{\rm nHe}}$
and $R_{12,{\rm eHe}}$ are radii of heavy MS stars and {\pbf
  enveloped He-burning} stars, respectively, and their ratio ($R_{\rm
  12, nHe}/R_{\rm 12, eHe}$) should be $\sim 0.01$. Then, we obtain
the collision-rate ratio such that
\begin{eqnarray}
    \frac{\Gamma_{\rm nHe}}{\Gamma_{\rm eHe}} &\sim& 10^{-2} \left(
    \frac{N_{1,{\rm nHe}}/N_{1,{\rm eHe}}}{2} \right)\left(
    \frac{M_{12,{\rm nHe}}/M_{12,{\rm eHe}}}{0.7} \right) \nonumber \\
    &\times& \left( \frac{R_{12,{\rm nHe}}/R_{12,{\rm eHe}}}{0.01}
    \right).
\end{eqnarray}
We therefore estimate that one of $\sim 100$ PI-gap BHs is formed
through collisions between heavy MS stars and naked He stars. This
estimate is consistent with the argument of \cite{2019arXiv191101434D}
that all the PI-gap BHs in their simulations are formed through
collisions between heavy MS and {\pbf enveloped He-burning}
stars. In their simulations, only $\lesssim 20$ PI-gap BHs are formed
in solar metallicity environment.

The solid red curve in Fig.\,1 of \cite{2019arXiv191101434D}, in which
the stellar mass drops just before collision, might indicate a
collision between heavy MS and naked He stars. This is because the He
star largely loses its envelope just before the collision. However, it
might be regarded as the {\pbf enveloped He-burning} case. In the
SSE code \citep{Hurley+2000}, {\pbf enveloped He-burning} stars
are assumed to have radii of $\sim 10^3R_\odot$, even if they lost
most of their envelopes. If it is the naked-He case, we can estimate
the fraction of BH via naked-He star to that via {\pbf enveloped
  He-burning} star, which corresponds to $\Gamma_{\rm nHe}/\Gamma_{\rm
  eHe}$ from the result of \citet{2019arXiv191101434D}. In
\citet{2019arXiv191101434D}, 1 of 6 PI-gap BHs involving BH-BH mergers
was possibly a naked-He star. This suggests that $\Gamma_{\rm
  nHe}/\Gamma_{\rm eHe} = 1/6 \sim 0.2$. Thus, we might underestimate
$\Gamma_{\rm nHe}/\Gamma_{\rm eHe}$ by more than $10$ times for
uncertain reasons, but we conservatively adopt $\Gamma_{\rm
  nHe}/\Gamma_{\rm eHe} \sim 10^{-2}$.

{\pbf We derive $P_{\rm b}$ as follows. We suppose that either of
  naked He stars and heavy MS stars should have companions since
  massive stars indicate multiplicities with a high probability
  \citep{2012Sci...337..444S}. We consider that a ratio of a primary
  mass to its companion mass is uniformly distributed in the range
  from $0$ to $1$, and that the semi-major axis distribution of
  binaries is flat in logarithmic scale from $10R_\odot$ to
  $10^5R_\odot$. We regard that LB-1-like systems have B-type stars
  with $2-25M_\odot$ and semi-major axes of $0.3-3$~au. Since the ZAMS
  masses of the naked He stars and heavy MS stars are $\sim
  50M_\odot$, we get $P_{\rm b} \sim 0.1$.}

Finally, we can calculate the formation rate of LB-1-like progenitors
from equation (\ref{eq:FormationRateOfLB-1-likeProgenitors}) as
\begin{eqnarray}
    \dot{N}_{\rm LB1,p} \sim {\pbf 3 \times 10^{-8} \left(
      \frac{P_{\rm b}}{0.1} \right) } \left( \frac{\rho_{\rm oc}}{10^4
      M_\odot \mbox{pc}^{-3}} \right) [\mbox{yr}^{-1}].
\end{eqnarray}
Here, we leave the factor of $\rho_{\rm oc}$, because the typical
initial density of open clusters is still uncertain and may be higher
than $10^4M_{\odot}\,\mbox{pc}^{-3}$
\citep{2010ARA&A..48..431P,2016ApJ...817....4F}.

\section{Tidal circularization}
\label{sec:TidalCircularization}

In previous section, we estimate the formation rate of LB-1-like
progenitors. The LB-1-like progenitors have high eccentricities due to
the collisions. They should be circularized, since the eccentricity of
the LB-1 system is $\sim 0.03$. A circularization mechanism is
necessary to explain the nearly zero eccentricity of the LB-1
system. Here, we evaluate the tidal circularization timescale after
the collision.

In the case of binary evolution, the most powerful circularization
process is tidal interaction. The efficiency of the tidal interaction
depends on the type of the stellar envelope. If the envelope is
convective, the tidal interaction is the equilibrium tide
\citep{1989A&A...220..112Z}. On the other hand, if the envelope is
radiative, the tidal interaction is the dynamical tide
\citep{1975A&A....41..329Z}. In order to determine the type of the
tidal effect, we consider the type of the envelopes of the collision
products. A collision product has a central He core made from a naked
He star with $20M_\odot$, and an H envelope made from a heavy MS star
with $50M_\odot$. Such a star has a convective envelope only when its
radius is $\gtrsim 10^3 R_\odot$, much larger than the semi-major axis
of the LB-1 system. Thus, only the dynamical tide with radiative
damping can circularize LB-1-like progenitors.

We use the circularization timescale of the dynamical tide with
radiative damping derived by \cite{1977A&A....57..383Z}, which is
\begin{eqnarray}
    \tau_{\rm cir,dyn} &=& \frac{2}{21} \left(\frac{GM_{\rm
        coll}}{R_{\rm coll}^3}\right)^{-\frac{1}{2}} \frac{M_{\rm
        coll}}{M_B} \nonumber \\
    &\times& \left(1+\frac{M_B}{M_{\rm
        coll}}\right)^{-\frac{11}{6}}E_2^{-1}\left(\frac{R_{\rm
        coll}}{a}\right)^{-\frac{21}{2}},
\end{eqnarray}
where $M_{\rm coll}$ and $R_{\rm coll}$ are the mass and radius of the
collision product, respectively, $a$ is the binary separation, and
$E_2$ is the second order tidal
coefficient. \cite{1975A&A....41..329Z} fitted $E_2$ as
\begin{equation}
    E_2 = 1.592 \times 10^{-9}M_{\rm coll}^{2.84}.
\end{equation}
We give $M_{\rm coll} = 70M_{\odot}$, $M_B = 8M_{\odot}$, and $a = 1$
au, which are the binary parameter of the LB-1 system, and calculate
the above equation as
\begin{equation}
    \tau_{\rm cir,dyn} \sim 5 \times 10^4 \left(\frac{R_{\rm
        coll}}{100R_\odot}\right)^{-9}{~\rm
      yr}. \label{eq:RadiativeCircularization}
\end{equation}
Since the lifetimes of the collision products are similar to that of
the naked He stars, $\sim 0.2$ Myr, the binaries should be soon
circularized if the collision products have radii of $\gtrsim
100R_\odot$. Note that the circularization timescale is $\sim 100$ yr
at $R_{\rm coll} \sim 200R_\odot$ due to the sharp dependence on the
radii of the collision products.

Although we do not know the initial radii of the collision products,
they should be larger than the radii of the colliding MS stars,
i.e. $\gtrsim 10R_\odot$. The collision products expand on
Kelvin-Helmholtz timescale:
\begin{eqnarray}
    t_{\rm KH} \lesssim 2 \times 10^5 \left( \frac{M_{\rm
        coll}}{70M_\odot} \right)^2 \left( \frac{R_{\rm
        coll}}{10R_\odot} \right)^{-1} \left( \frac{L_{\rm coll}}{10^5
      L_\odot} \right)^{-1} [\mbox{yr}].
\end{eqnarray}
Since the lifetime of the collision product is $\sim 0.2$ Myr, the
collision product should expand to $\gtrsim 100R_\odot$.

When the radius of the collision products exceed $\sim 200R_\odot$ (or
$\sim 1$ au), the binaries start Roche-lobe overflow. Since the mass
ratios of the collision products to the B-type stars are about $10$,
the separations of the binaries become small so quickly that the
Roche-lobe overflow becomes unstable. Such binaries should undergo
common envelope evolution. The common envelope evolution blows the
envelopes of the collision products, and the collision products
collapse to BHs as massive as the pre-existing naked He stars, not
PI-gap BHs. Thus, if the collision products expand to $\gtrsim
200R_\odot$, the LB-1-like progenitors cannot become LB-1-like
systems. Since $t_{\rm KH} \sim 2 \times 10^4$ yr at $R_{\rm coll}
\sim 100R_\odot$, and the lifetime of the collision products are $0.2$
Myr, the probability that the collision products collapse to PI-gap
BHs at $R_{\rm coll} \sim 100$-- $200R_\odot$ is estimated to be
$10$\,\%. Note that the collapse time of the collision products is at
random during their lifetimes, since their He cores are originally
naked He stars, and are wandering a long time before their collisions.

Combining this with the results of the previous section, we finally
obtain the number of LB-1-like systems in the MW galaxy as
\begin{eqnarray}
    N_{\rm LB1} \sim \left\{
    \begin{array}{ll}
    {\pbf 0.01 (P_{\rm b}/0.1)} (\rho_{\rm oc}/10^4M_\odot
    \mbox{pc}^{-3}) \\ (T_{\rm B}/40\mbox{Myr}) & [M_{\rm B} \gtrsim
      8M_\odot] \\
    & \\
    {\pbf 0.3 (P_{\rm b}/0.1)} (\rho_{\rm oc}/10^4M_\odot
    \mbox{pc}^{-3}) \\ (T_{\rm B}/1\mbox{Gyr}) & [M_{\rm B} \gtrsim
      2M_\odot]
    \end{array}
    \right..
\end{eqnarray}
The lifetime of LB-1-like systems is the lifetime of B-type stars,
$T_{\rm B}$. After B-type stars end their evolution, LB-1-like systems
cannot be observed. We consider a LB-1-like progenitor consisting of a
collision product and O-type star with $\sim 25M_\odot$, not a B-type
star. When the radius of the collision product exceeds $\sim
200R_\odot$, stable Roche-lobe overflow (not common envelope
evolution) starts, since the collision product has a radiative
envelope, and the mass ratio of the collision product to the O-type
star is less than three. Then, the binary always survive and is
circularized, although only $10$ \% of a binary survives for the case
of a B-type companion star. If the Roche-lobe overflow {\pbf does
  not reduce the mass} of the collision product, the collision product
collapses to a PI-gap BH. Then, a LB-1-like system appears, although
the system has an O-type star. Since the O-type star has a lifetime of
$\sim 7$ Myr, we can estimate the number of such LB-1-like systems in
the MW galaxy as
\begin{eqnarray}
    N_{\rm LB1} &\sim& {\pbf 0.02 (P_{\rm b}/0.1)} (\rho_{\rm
      oc}/10^4M_\odot \mbox{pc}^{-3}) \nonumber \\
    &\times& (T_{\rm O}/7\mbox{Myr}) \;\; [M_{\rm O} \gtrsim 25M_\odot],
\end{eqnarray}
where $T_{\rm O}$ and $M_{\rm O}$ are the lifetime and mass of an
O-type star. {\pbf Note that $P_{\rm b}$ is similar to that in
  the case of B-type companions.} Finally, this mechanism is not
efficient enough to explain the presence of the LB-1 system, even if
we assume companion stars of PI-gap BHs to be O-type stars.

\section{Other possible scenarios}
\label{sec:OtherPossibleScenarios}

\subsection{Stellar collisions}

We can divide stellar types into three: MS stars, {\pbf enveloped
  He-burning} stars, and naked He stars. Collisions with {\pbf
  enveloped He-burning} stars do not work. Such stars have radii of
$\gtrsim 10^3R_\odot$, more than $1$ au. The collision products
swallow companion stars when the binary separations are $\sim 1$
au. We consider collisions between two MS stars, and between two naked
He stars. Their total masses should exceed $70M_\odot$. Then, they
cannot avoid PPISNe/PISNe. Thus, such collisions do not work for the
formation of the LB-1 system. Finally, collisions between MS and naked
He stars have a event rate not enough to explain the formation of the
LB-1 system described in the previous sections. In summary, any types
of stellar collisions cannot explain the presence of the LB-1 system.

\subsection{Capture scenarios}

As seen in section \ref{sec:FormationRateOfLB-1-likeProgenitors}, many
PI-gap BHs are formed in open clusters. In this section, we consider
whether they can capture B-type stars, and whether they can form
LB-1-like systems.

\subsubsection{Open clusters}
\label{sec:OpenCluster}

We first estimate the number of binaries with PI-gap BHs and B-type
stars, $N_{\rm b}$. It is expressed as
\begin{eqnarray}
    N_{\rm b} = \Gamma_{\rm cap} T_{\rm
      B}, \label{eq:NumberOfLB-1Candidate}
\end{eqnarray}
where $\Gamma_{\rm cap}$ is a rate at which PI-gap BHs capture B-type
stars. We calculate $\Gamma_{\rm cap}$, supposing this capture
mechanism is encounters between PI-gap BHs and binaries with B-type
stars. We calculate this rate, using equations (\ref{eq:AnyEventRate})
and (\ref{eq:SweepingVolume}). In this case, $N_1$ is the number of
{\pbf PI-gap BHs during the period where B-type stars are on the
  MS}, which can be given by
\begin{eqnarray}
    N_1 = \dot{N}_{\rm PIgap} T_{\rm B} \sim 80 \left( \frac{\rho_{\rm
        oc}}{10^4 M_\odot\mbox{pc}^{-3}} \right) \left( \frac{T_{\rm
        B}}{40\mbox{Myr}} \right),
\end{eqnarray}
where we adopt the value in equation (\ref{eq:FormationRateOfPiGap})
for $\dot{N}_{\rm PIgap}$, and the lifetime of $8M_\odot$ stars for
$T_{\rm B}$. For $n_2$, we choose the number density of MS stars with
more than $8M_\odot$. Then, $n_2$ can be calculated as
\begin{eqnarray}
    n_{2} &=& \eta_8 \rho_{\rm oc} \nonumber \\
    &\sim& 7 \times 10^{-55} \left( \frac{\eta_{\rm
        B}}{0.02M_\odot^{-1}} \right) \left( \frac{\rho_{\rm
        oc,late}}{10^3 M_\odot\mbox{pc}^{-3}} \right) [\mbox{cm}^3],
\end{eqnarray}
where $\rho_{\rm oc,late}$ is mass density of open clusters after
PI-gap BHs are formed. Since open clusters have lost large amounts of
mass by that time, $\rho_{\rm oc,late}$ should be much less than
$\rho_{\rm oc}$. The sweeping volume $V_{12}$ should be
\begin{eqnarray}
    V_{12} &\sim& 1 \times 10^{37} \left( \frac{M_{12}}{100M_\odot}
    \right) \nonumber \\
    &\times& \left( \frac{R_{12}}{1\mbox{au}} \right) \left(
    \frac{v_{12}}{1\mbox{kms}^{-1}} \right)^{-1} [\mbox{cm}^3
      \mbox{s}^{-1}],
\end{eqnarray}
where we adopt PI-gap BH mass for $M_{12}$, semi-major axes of the
LB-1 system for $R_{12}$, and velocity dispersion in open clusters for
$v_{12}$. Finally, we get $N_{\rm b}$ as
\begin{eqnarray}
    N_{\rm b} \sim \left\{
    \begin{array}{ll}
      0.7 \left( \rho_{\rm oc}/10^4 M_\odot\mbox{pc}^{-3} \right)
      \\ (T_{\rm B}/40\mbox{Myr}) & [M_{\rm B} \gtrsim 8M_\odot] \\
      & \\
      20 \left( \rho_{\rm oc}/10^4 M_\odot\mbox{pc}^{-3} \right)
      \\ (T_{\rm B}/1\mbox{Gyr}) & [M_{\rm B} \gtrsim 2M_\odot]
    \end{array}
    \right..
\end{eqnarray}

These binaries are eccentric LB-1-like systems due to the
binary-single encounters, and however they cannot be circularized
after their formation as follows. The B-type stars are MS stars, and
have radiative envelopes. Since their radii are $\lesssim 10R_\odot$,
the circularization timescale is $\tau_{\rm cir,dyn} \sim 5 \times
10^{13}$ yr according to equation
(\ref{eq:RadiativeCircularization}). This is much larger than the
lifetimes of the B-type stars. Thus, their eccentricities keep
constant from their formation time to their ending time.

We estimate the probability that these binaries have nearly zero
eccentricities, say $\lesssim 0.05$. We assume that binary-single
encounters leave binaries with eccentricities in the thermal
distribution \citep{1975MNRAS.173..729H}. The probability of these
binaries with eccentricities of $\lesssim 0.05$ is $\sim
10^{-3}$. Thus, this mechanism does not work for the formation of the
LB-1 system.

The above number of eccentric LB-1-like systems put strong constraints
on any other capture scenarios in open clusters. This is because the
above scenario has the highest event rate among the capture scenarios
in open clusters. For example, let's consider collision of a BH with a
MS star whose companion star is a B-type star. {\pbf Note that
  this collision in dense stellar clusters is intensively investigated
  by \cite{2019arXiv191206022B}.}  We assume that a binary consisting
of the collision product and B-type star receives tidal friction
through unknown processes much more efficiently than we consider in
section \ref{sec:TidalCircularization}, and that a LB-1-like system is
formed. However, the collision rate is $\sim 10$ times less than the
capture rate of B-type stars by PI-gap BHs discussed above, since the
radius of a MS star is $\sim 10R_\odot$ less than the semi-major axis
of the LB-1 system ($\sim 1$ au) by $\sim 10$. Thus, before the
discovery of LB-1-like systems formed through this mechanism, we
should {\pbf have discovered} eccentric LB-1-like systems formed
through capture of B-type stars by PI-gap BHs.  Although radial
velocity observations can discover circular binaries more easily than
eccentric binaries, the detection efficiency for circular binaries is
more than for eccentric binaries only by $10$ \%
\citep{2008ApJ...685..553S}. In summary, any capture scenarios do not
work for the formation of the LB-1 system.

\subsubsection{Globular clusters}

Many PI-gap BHs should be formed in globular clusters. We consider
whether they can form LB-1-like systems. Since globular clusters are
old in the MW galaxy ($\gtrsim 10$~Gyr), their turnoff masses should
be $\sim 0.8M_\odot$. Thus, even if turnoff stars merge, and form blue
stragglers, their masses should be $\sim 1.6M_\odot$. They are not
B-type stars. In summary, PI-gap BHs in globular clusters cannot
capture B-type stars, since there is no B-type star in globular
clusters. Thus, LB-1-like systems cannot be formed in globular
clusters.

\subsubsection{Interstellar space}

Pop. II/III stars form a large number of merging BH-BHs with the total
masses of $\sim 70M_\odot$, and form PI-gap BHs
\citep[e.g.][]{2014MNRAS.442.2963K, 2016Natur.534..512B}. Thus, many
PI-gap BHs should be wandering in the MW galaxy. We assess whether
they can capture B-type stars in the MW galaxy. We suppose this
mechanism is binary-single encounters as the same in section
\ref{sec:OpenCluster}.

The number of such binaries can be written in the same way as
equations (\ref{eq:NumberOfLB-1Candidate}). However, we should
calculate $\Gamma_{\rm cap}$ in a different way. We adopt the number
of Pop. III PI-gap BHs for $N_1$, and estimate it from
\cite{2014MNRAS.442.2963K} as
\begin{eqnarray}
    N_1 \sim 10^5.
\end{eqnarray}
We calculate $n_2$ as the number density of B-type stars in the MW
galaxy in the following:
\begin{eqnarray}
    n_2 &=& \frac{\eta_8 \dot{M}_{\rm mw} T_{\rm B}}{4/3 \pi R_{\rm
        mw}^3} \nonumber \\
    &\sim& 10^{-62} \left( \frac{\eta_8}{0.01M_\odot^{-1}} \right)
    \left( \frac{\dot{M}_{\rm mw}}{2M_\odot\mbox{yr}^{-1}} \right) \nonumber \\
    &\times& \left( \frac{T_{\rm B}}{40\mbox{Myr}} \right) \left(
    \frac{R_{\rm mw}}{10\mbox{kpc}} \right)^{-3} [\mbox{cm}^{-3}].
\end{eqnarray}
where $R_{\rm mw}$ is the size of the MW galaxy. Although B-type stars
are formed in the MW disk, Pop. III PI-gap BHs should be wandering in
the MW halo. Thus, we assume B-type stars spread in the MW halo. The
sweeping volume is given by
\begin{eqnarray}
    V_{12} &\sim& 6 \times 10^{34} \left( \frac{M_{\rm 12}}{100M_\odot}
    \right) \nonumber \\
    &\times& \left( \frac{R_{12}}{1\mbox{au}} \right) \left(
    \frac{v}{200\mbox{kms}^{-1}} \right)^{-1} [\mbox{cm}^3
      \mbox{s}^{-1}],
\end{eqnarray}
where we adopt the MW circular velocity for $v_{12}$. Putting together
the above equations, we obtain the capture rate as
\begin{eqnarray}
    N_{\rm bin} \sim 7 \times 10^{-8} [M_{\rm B} \gtrsim 8M_\odot].
\end{eqnarray}
If we also take into account Pop. II PI-gap BHs, the number should be
increased slightly.

Any other capture scenarios in interstellar space do not work. The
above event rate is the highest among any capture scenarios in
interstellar space, similarly to the discussion in section
\ref{sec:OpenCluster}.

\subsection{Hierarchical triple system}

We explore the possibility that LB-1-like systems are formed from
hierarchical triple systems. We consider a hierarchical triple system
which has an inner binary consisting of heavy MS stars, and a B-type
star as the third star. Since the inner binary has to leave a PI-gap
BH through merger/collision, we can assume their total mass to be
$\sim 100M_\odot$ and the third star to be $\sim 10M_\odot$. We can
also assume that the outer binary has semi-major axis of $a_{\rm out}
\sim 1$ au, while the inner binary has semi-major axis of $a_{\rm in}
\sim 10-100R_\odot$. If $a_{\rm in} < 10R_\odot$, the inner binary
merges when they are MS stars. If $a_{\rm out} > 100R_\odot$, the
hierarchical triple system is unstable
\citep{1972CeMec...6..322H,1999ASIC..522..385M}.

The inner binary cannot leave a PI-gap BH through pure binary
evolution for the following reason. In order to produce a PI-gap BH,
they have to merge when the primary star is an {\pbf enveloped
  He-burning} star, and the secondary star is an MS star. {\pbf
  This merger can be driven by two mechanisms: common envelope
  evolution and Case B merger
  \citep{1994A&A...290..119P,2001A&A...369..939W,2010NewAR..54...39P,2014ApJ...796..121J}.}
First, we assess whether the inner binary satisfies conditions of the
onset criteria of common envelope evolution. Since $a_{\rm in}
\lesssim 100R_\odot$, the primary star has a radiative envelope when
it begins interacting with the secondary star. Therefore, the mass
ratio of the primary star to the secondary star should be large (more
than three). In order to produce a LB-1-like system, on the other
hand, the secondary star should have $\gtrsim 50M_\odot$, since the
common envelope evolution blows away the primary envelopes and leaves
the He core of the primary star with a mass of $\sim 20M_\odot$. Thus,
the primary star initially should have $\gtrsim 150M_\odot$ because in
the case of solar metallicity, stellar wind halves the initial mass by
the end of the MS phase. However, if the primary star has
$>150M_\odot$, its radius must exceed $100R_\odot$ before the end of
the MS phase, and it merges with the secondary star. Since both of the
primary and secondary stars are MS stars, their merger remnant is also
a MS star. It makes a large He core, and cannot leave a PI-gap BH due
to PPISN/PISN.

{\pbf Second, we assess Case B merger in which a star in a
  Hertzsprung gap phase merges with an MS star. In order to form a
  $70M_\odot$ BH from the merger product, the merger product and
  primary star have at least $70M_\odot$ and $35M_\odot$,
  respectively. We estimate a radius of a merger product at the
  evolutionary endpoint, based on Fig.~3 of \cite{2014ApJ...796..121J}
  who have investigated the evolution of merger products with primary
  masses of $20$, $25$, and $30M_\odot$. The figure shows two
  points. First, a merger product with a larger primary mass has a
  larger radius at the evolutionary endpoint if we fix the mass of the
  merger product. Second, a merger product with a larger mass has a
  larger radius at the evolutionary endpoint if the merger product is
  made from two equal-mass stars. Thus, a merger product made from two
  $35M_\odot$ stars has the smallest radius at the evolutionary
  endpoint among merger products which can form $70M_\odot$ BHs. We
  obtain its radius as follows. Its luminosity should be $\sim
  10^{5.9}L_\odot$, since luminosities of merger products with $\sim
  70M_\odot$ are $10^{5.9}L_\odot$ regardless of their primary
  masses. Its effective temperature should be $\sim 10^{4.05}$~K,
  since merger products with $\sim 70M_\odot$ have effective
  temperatures of $10^{4.4}$, $10^{4.3}$, and $10^{4.15}$~K for the
  primary masses of $20$, $25$, and $30M_\odot$, respectively. Thus,
  the merger product has a radius of $\gtrsim 1$~au at the
  evolutionary endpoint. Since the merger product and third star are
  separated only by $\sim 1$~au, they experience mass transfer from
  the merger product to the third star. This mass transfer is unstable
  (i.e. common envelope evolution), because the mass ratio of the
  merger product to the third star is high ($\gtrsim 10$). Then, the
  merger product loses its envelope, and cannot leave a $70M_\odot$
  BH. Finally, we conclude that Case B merger cannot form a LB-1-like
  system.}

The inner MS-MS binary may collide through secular interaction between
the outer binary before either of them evolves to an {\pbf
  enveloped He-burning} star. Here, we take into account Kozai-Lidov
(KL) mechanism as secular interaction \citep{1962AJ.....67..591K}. The
KL timescale can be expressed as
\begin{eqnarray}
    T_{\rm KL} = 2\pi \frac{\left(Gm_{\rm
        in}\right)^{1/2}}{Gm_3}\frac{a_{\rm out}^3}{a_{\rm in}^{3/2}}
    \left( 1 - e_{\rm out}^2\right),
\end{eqnarray}
where $m_{\rm in}$ and $m_3$ are the masses of the inner binary and
third star, and $e_{\rm out}$ is the eccentricity of the outer
binary. This can be calculated as
\begin{eqnarray}
    T_{\rm KL} &\lesssim& 100 \left( \frac{m_{\rm in}}{100M_\odot}
    \right)^{1/2} \left( \frac{m_3}{10M_\odot} \right)^{-1} \nonumber \\
    &\times& \left( \frac{a_{\rm in}}{10R_\odot} \right)^{-3/2} \left(
    \frac{a_{\rm out}}{1\mbox{au}} \right)^3 [\mbox{yr}],
\end{eqnarray}
where the equal sign of the above equation is held for $e_{\rm
  out}=0$. Therefore, if KL mechanism works, the inner binary merges
before the primary star evolves to an {\pbf enveloped He-burning}
star. From the above discussion, we conclude that hierarchical triple
systems cannot form LB-1-like binaries.

\section{Summary}
\label{sec:Summary}

We assess various mechanisms forming LB-1-like systems through
dynamical interactions, not assuming the reduction of stellar wind
mass loss. The most efficient mechanism is collision of naked He stars
with heavy MS stars which have B-type companion stars in open
clusters. The number of LB-1-like systems formed through this
mechanism is estimated to be $\sim {\pbf 0.01 (P_{\rm b}/0.1)}
(\rho_{\rm oc}/10^4M_\odot\mbox{pc}^{-3})$ in the MW galaxy. If we
take into account LB-1-like systems with O-type stars as companion
stars, the number increases to $\sim {\pbf 0.03 (P_{\rm b}/0.1)}
(\rho_{\rm oc}/10^4M_\odot\mbox{pc}^{-3})$.

This mechanism can form LB-1-like systems at least {\pbf 10}
times more efficiently than any other mechanisms: capture of B-type
stars by PI-gap BHs, stellar collisions between other type stars, and
stellar mergers in hierarchical triple systems. Especially, capture
scenarios result in too many eccentric binaries. If one of the capture
scenarios formed the LB-1 system, we would have detected eccentric
LB-1-like systems earlier than the LB-1 system.

\begin{ack}
We thank T. Yoshida and Y. Hori for fruitful advice. AT is grateful
for the hospitality of Nicolaus Copernicus Astronomical Center during
a research visit at which this work was initiated. MF was supported by
The University of Tokyo Excellent Young Researcher Program.  This work
was supported by JSPS KAKENHI Grant Number 17H06360, 18J00558,
19H01933, and 19K03907.
\end{ack}


\end{document}